\documentclass[12pt,a4wide,epsf]{article}
\usepackage{epsf}

\usepackage{graphicx}

\newcommand{\insertplot}[5]{\begin{figure}
 \hfill\hbox to 0.05in{\vbox to #5in{\vfill
 \inputplot{#1}{#4}{#5}}\hfill}
 \hfill\vspace{-.1in}
 \caption{#2}\label{#3}
 \end{figure}}
 \newcommand{\inputplot}[3]{
 \special{ps: plotfile #1}
}
\newcounter{fig}   

\newcommand{\vphi}{\varphi}
\newcommand{\vepsilon}{\varepsilon}

\newcommand{\sqdetg}{\sqrt{-g}}

\usepackage{graphicx}

\begin{document}

\title{
Negative Horizon Mass for Rotating Black Holes}
 \vspace{1.5truecm}
\author{
{\bf Jutta Kunz} and 
{\bf Francisco Navarro-L\'erida} \\
Institut f\"ur Physik, Universit\"at Oldenburg, Postfach 2503\\
D-26111 Oldenburg, Germany\\
}

\vspace{1.5truecm}

\date{\today}

\maketitle
\vspace{1.0truecm}

\begin{abstract}
Charged rotating black holes of Einstein-Maxwell-Chern-Simons theory 
in odd dimensions, $D \ge 5$, may possess a negative horizon mass,
while their total mass is positive. This surprising feature is related
to the existence of counterrotating solutions, 
where the horizon angular velocity $\Omega$
and the angular momentum $J$ possess opposite signs. 
Black holes may further possess vanishing horizon angular velocity 
while they have finite angular momentum, or they may possess
finite horizon angular velocity while their angular momentum vanishes.
In $D=9$ even non-static black holes with $\Omega=J=0$ appear.
Charged rotating black holes with vanishing gyromagnetic ratio exist, and
black holes need no longer be uniquely characterized by their global charges. 
\end{abstract}

\vfill\eject

\section{Introduction}

The Kerr-Newman (KN) family of black hole solutions of Einstein-Maxwell (EM) theory
\cite{K}
possesses many special properties, not present in black hole solutions of theories
with more general matter content. The inclusion of a dilaton, for instance,
leads to deviations of the gyromagnetic ratio from the KN value \cite{dil},
and when the dilaton coupling exceeds the Kaluza-Klein value
counterrotating black holes appear, where the
horizon rotates in the opposite sense to the total angular momentum \cite{KKN};
while the EM uniqueness theorem or Israel's theorem \cite{unique}
do not hold in general in theories with non-Abelian fields \cite{eym1,eym2}.

In the context of supergravity or string theory, higher dimensional black holes 
are of interest. The Myers-Perry solutions \cite{MP}, representing 
higher dimensional rotating vacuum black holes, have long been known,
as well as various charged rotating black holes of supergravity and string theory
\cite{youm}. In contrast, the higher dimensional generalization of the KN black holes
has so far resisted attempts to obtain them in analytical form \cite{KNP, EM_KNV}. 

In odd dimensions, $D=2N+1$, the EM action may be supplemented by an `$A\,F^N$'
Chern-Simons (CS) term. In 5 dimensions, for a certain value of the CS 
coefficient $\lambda=\lambda_{\rm SG}$, 
the theory corresponds to the bosonic sector of $D=5$ supergravity,
where rotating black hole solutions are known analytically \cite{BLMPSV,Cvetic}.
In particular, extremal solutions exist,
whose horizon angular velocity vanishes.
Thus their horizon is non-rotating, although their angular momentum is nonzero.
In these solutions a negative fraction of the total angular momentum
is stored in the Maxwell field behind the horizon,
and the effect of rotation on the horizon is not to make it rotate
but to deform it into a squashed 3-sphere \cite{GMT,surprise}.
Here supersymmetry is associated with a borderline between stability and instability,
since for $\lambda>\lambda_{\rm SG}$ a rotational instability arises,
where counterrotating black holes appear \cite{KN}.
Moreover, when the CS coefficient is increased beyond the $2\lambda_{\rm SG}$,
black holes (with horizon topology of a sphere \cite{blackrings})
are no longer uniquely characterized by their global charges \cite{KN}.

Here we reanalyze 5-dimensional Einstein-Maxwell-Chern-Simons
(EMCS) black holes and show, that they may possess
a negative horizon mass for a CS coefficient beyond $2\lambda_{\rm SG}$.
We also investigate their higher dimensional generalizations
for arbitrary CS coefficient. 
%
While $D=2N+1$-dimensional black holes generically possess
$N$ independent angular momenta \cite{MP},
we here focus on black holes with equal-magnitude angular momenta \cite{EM_KNV},
allowing for the reduction of the EMCS equations to a set of ordinary differential
equations. In section 2 we recall the EMCS action and
present the appropriate Ans\"atze for the metric and the gauge potential.
We discuss the black hole properties in section 3. 
We present the numerical results for $D=5$ and $D>5$ black holes
in section 4, and summarize in section 5.

\section{Action, Metric and Gauge Potential}

We consider the odd $D$-dimensional EMCS
action with Lagrangian \cite{GMT}
\begin{equation}
L = \frac{1}{16 \pi G_D} \sqdetg  \left(R - F_{\mu \nu} F^{\mu \nu} +
\frac{8}{D+1}{\tilde \lambda} \epsilon^{\mu_1 \mu_2 \dots
  \mu_{D-2}\mu_{D-1}\mu_D} F_{\mu_1 \mu_2} \dots F_{\mu_{D-2} \mu_{D-1}} A_{\mu_D}    \right) \ , \label{action}
\end{equation}
with curvature scalar $R$,
$D$-dimensional Newton constant $G_D$,
and field strength tensor
$
F_{\mu \nu} =
\partial_\mu A_\nu -\partial_\nu A_\mu $,
where $A_\mu $ denotes the gauge potential. ${\tilde \lambda}$ corresponds to
the CS coupling constant.

The field equations are obtained from the action Eq.~(\ref{action}) 
by taking variations
with respect to the metric and the gauge potential,
yielding the Einstein equations
\begin{equation}
G_{\mu\nu}= R_{\mu\nu}-\frac{1}{2}g_{\mu\nu}R = 2 T_{\mu\nu}
\ , \label{ee}
\end{equation}
with stress-energy tensor
\begin{equation}
T_{\mu \nu} = F_{\mu\rho} {F_\nu}^\rho - \frac{1}{4} g_{\mu \nu} F_{\rho
  \sigma} F^{\rho \sigma} \ ,
\end{equation}
and the gauge field equations
\begin{equation}
\nabla_\nu F^{\mu_1 \nu}  =  {\tilde \lambda}  \epsilon^{\mu_1 \mu_2 \mu_3 \dots
  \mu_{D-1}\mu_D} F_{\mu_2 \mu_3} \dots F_{\mu_{D-1} \mu_D}   \ .
\label{feqA}
\end{equation}

In order to obtain stationary black hole solutions of Eqs.~(\ref{ee})-(\ref{feqA}),
representing charged generalizations of the odd D-dimensional
Myers-Perry solutions \cite{MP},
we consider black hole space-times with $N$-azimuthal symmetries,
implying the existence of $N+1$ commuting Killing vectors,
$\xi \equiv \partial_t$, 
and $\eta_{(k)} \equiv \partial_{\vphi_k}$, for $k=1, \dots , N$, where $N$ is
defined by $D=2 N + 1$.

While generic EMCS black holes possess $N$ independent
angular momenta, we now restrict to black holes whose 
angular momenta have all equal magnitude.
The metric and the gauge field parametrization then
simplify considerably. 
In particular, for such equal-magnitude angular momenta
black holes, the general Einstein and gauge field equations reduce to
a set of ordinary differential equations \cite{EM_KNV}, since the angular
dependence can be treated explicitly.

Restricting to black holes with spherical horizon topology,
we parametrize the metric in isotropic coordinates,
which are well suited for the numerical construction of
rotating black holes \cite{KNP,KN,kkrot}, 
\begin{eqnarray}
&&ds^2 = -f dt^2 + \frac{m}{f} \left[ dr^2 + r^2 \sum_{i=1}^{N-1}
  \left(\prod_{j=0}^{i-1} \cos^2\theta_j \right) d\theta_i^2\right] \nonumber \\
&&+\frac{n}{f} r^2 \sum_{k=1}^N \left( \prod_{l=0}^{k-1} \cos^2 \theta_l
  \right) \sin^2\theta_k \left(\vepsilon_k d\vphi_k - \frac{\omega}{r}
  dt\right)^2 \nonumber \\
&&+\frac{m-n}{f} r^2 \left\{ \sum_{k=1}^N \left( \prod_{l=0}^{k-1} \cos^2
  \theta_l \right) \sin^2\theta_k  d\vphi_k^2 \right. \nonumber\\
&& -\left. \left[\sum_{k=1}^N \left( \prod_{l=0}^{k-1} \cos^2
  \theta_l \right) \sin^2\theta_k \vepsilon_k d\vphi_k\right]^2 \right\} \ ,
  \label{metric}
\end{eqnarray}
where $\theta_0 \equiv 0$, $\theta_i \in [0,\pi/2]$ 
for $i=1,\dots , N-1$, 
$\theta_N \equiv \pi/2$, $\vphi_k \in [0,2\pi]$ for $k=1,\dots , N$,
and $\vepsilon_k = \pm 1$ denotes the sense of rotation
in the $k$-th orthogonal plane of rotation.

The parametrization for the gauge potential, consistent with
Eq. (\ref{metric}), is
\begin{equation}
A_\mu dx^\mu =  a_0 dt + a_\vphi \sum_{k=1}^N \left(\prod_{l=0}^{k-1}
  \cos^2\theta_l\right) \sin^2\theta_k \vepsilon_k d\vphi_k \ .
\label{gaugepotential}
\end{equation}
 
It is remarkable that, independent of the odd dimension $D\ge 5$,
this parametrization involves only four functions $f, m, n, \omega$
for the metric and two functions $a_0, a_\vphi$
for the gauge field, which all depend only on the radial coordinate $r$.

\section{Black Hole Properties}

\subsection{Boundary conditions}

By substituting Eqs.~(\ref{metric})-(\ref{gaugepotential}) in
Eqs.~(\ref{ee})-(\ref{feqA}), a system of 6 ordinary differential equations is
obtained. In order to generate black hole solutions, appropriate boundary
conditions have to be imposed. 

For asymptotically flat solutions, the metric functions should satisfy
at infinity the boundary conditions
\begin{equation}
f|_{r=\infty}=m|_{r=\infty}=n|_{r=\infty}=1 \ , \ \omega|_{r=\infty}=0 \ .
\label{bc1} \end{equation}
For the gauge potential we choose a gauge, in which it vanishes
at infinity
\begin{equation}
a_0|_{r=\infty}=a_\vphi|_{r=\infty}=0 \ .
\label{bc2} \end{equation}

The horizon is located at $r=r_{\rm H}$,
and is characterized by the condition $f(r_{\rm H})=0$ \cite{kkrot}.
Requiring the horizon to be regular, the metric functions must
satisfy the boundary conditions
\begin{equation}
f|_{r=r_{\rm H}}=m|_{r=r_{\rm H}}=n|_{r=r_{\rm H}}=0 \ ,
\ \omega|_{r=r_{\rm H}}=r_{\rm H} \Omega \ , 
\label{bc3} \end{equation}
where $\Omega$ is (related to) the horizon angular velocity, 
defined in terms of the Killing vector
\begin{equation}
\chi = \xi + \Omega \sum_{k=1}^N \vepsilon_k \eta_{(k)} \ ,
\label{chi} \end{equation}
which is null at the horizon. The gauge potential satisfies
\begin{equation}
\left. \chi^\mu A_\mu \right|_{r=r_{\rm H}} =
\Phi_{\rm H} = \left. (a_0+\Omega a_\vphi)\right|_{r=r_{\rm H}} \ , \ \ \
\left. \frac{d a_\vphi}{d r}\right|_{r=r_{\rm H}}=0
\ , \label{bc4} \end{equation}
with constant horizon electrostatic potential $\Phi_{\rm H}$.

\subsection{Global charges}

Since the space-times we are considering are stationary,
(multi-)axisymmetric, and asymptotically flat we may compute the mass $M$ and
the $N$ angular momenta $J_{(k)}$ of the black holes by means of the Komar expressions associated with the respective Killing vector fields
\begin{equation}
M = \frac{-1}{16 \pi G_D} \frac{D-2}{D-3} \int_{S_{\infty}^{D-2}} \alpha \ , \ \ \
J_{(k)} = \frac{1}{16 \pi G_D}  \int_{S_{\infty}^{D-2}} \beta_{(k)} \ , \label{Komar_MJ}
\end{equation}
with $\alpha_{\mu_1 \dots \mu_{D-2}} \equiv \epsilon_{\mu_1 \dots \mu_{D-2}
  \rho \sigma} \nabla^\rho \xi^\sigma$,
$\beta_{ (k) \mu_1 \dots \mu_{D-2}} \equiv \epsilon_{\mu_1 \dots \mu_{D-2}
  \rho \sigma} \nabla^\rho \eta_{(k)}^\sigma$.
For equal-magnitude angular momenta $J_{(k)}=\vepsilon_k J$, 
$k=1, \dots , N$.

The electric charge $Q$ associated with the Maxwell field can be defined by
\begin{equation}
Q=\frac{-1}{8 \pi G_D} \int_{S_{\infty}^{D-2}} \tilde F \ , \label{elect_charge}\end{equation}
with ${\tilde F}_{\mu_1 \dots \mu_{D-2}} \equiv  \epsilon_{\mu_1 \dots \mu_{D-2} \rho \sigma} F^{\rho \sigma}$.

These global charges and the magnetic moment $\mu_{\rm mag}$
can be obtained from the asymptotic expansions of the metric and the gauge
potential 
\begin{eqnarray}
f=1-\frac{\hat M}{r^{D-3}} + \dots \ ,  \ \ \
\omega=\frac{\hat J}{r^{D-2}}  + \dots \ ,  \ \ \
a_0=\frac{\hat Q}{r^{D-3}} + \dots \ ,  \ \ \
a_\vphi=-\frac{{\hat \mu}_{\rm mag}}{r^{D-3}} + \dots \ ,
\end{eqnarray}
where
\begin{eqnarray}
{\hat M}=\frac{16 \pi G_D}{(D-2)A(S^{D-2})} M \ &,& \ \ \
{\hat J}=\frac{8 \pi G_D}{A(S^{D-2})}J \ , \nonumber \\
{\hat Q}=\frac{4 \pi G_D}{(D-3)A(S^{D-2})} Q \ &,& \ \ \
{\hat \mu}_{\rm  mag}=\frac{4 \pi G_D}{(D-3)A(S^{D-2})} \mu_{\rm mag} \ ,
\end{eqnarray}
and $A(S^{D-2})$ denotes the area of the unit $(D-2)$-sphere.
The gyromagnetic ratio $g$ is then defined via
\begin{equation}
g=2\frac{M {\mu_{\rm mag}}}{Q J} \ . \label{g_factor}
\  \end{equation}

\subsection{Mass formulae}

The surface gravity $\kappa$ of the black holes is defined by
\begin{equation}
\kappa^2 = -\frac{1}{2} \lim_{r \to r_{\rm H}} (\nabla_\mu \chi_\nu)
  (\nabla^\mu \chi^\nu) \ . \label{surface_grav}
\end{equation}
For equal-magnitude angular momenta,
the area of the horizon $A_{\rm H}$ reduces to
\begin{equation}
A_{\rm H}= r_{\rm H}^{D-2} A(S^{D-2}) \lim_{r \to r_{\rm H}}
\sqrt{\frac{m^{D-3} n}{f^{D-2}}} \ . \label{hor_area}
\end{equation}

The horizon mass $M_{\rm H}$ and horizon angular momenta
$J_{{\rm H} (k)}$ are given by
\begin{equation}
M_{\rm H} = \frac{-1}{16 \pi G_D} \frac{D-2}{D-3} \int_{{\cal H}} \alpha \ , \ \ \
J_{{\rm H} (k)} = \frac{1}{16 \pi G_D}  \int_{{\cal H}} \beta_{(k)} \ , \label{hor_MJ}
\end{equation}
where ${\cal H}$ represents the surface of the horizon,
and for equal-magnitude angular momenta
$J_{{\rm H} (k)} =\vepsilon_k J_{\rm H}$, $k=1, \dots , N$.
The mass $M$ and angular momenta $J_{(k)}$ may thus be reexpressed in the
form \cite{GMT,surprise}
\begin{equation}
 M=M_{\rm H} + M_{\Sigma} \ , \ \ \
 J_{(k)}=J_{{\rm H} (k)} + J_{\Sigma (k)} \ , \label{sum_MJ}
\end{equation}
where $J_{\Sigma (k)}$ is a `bulk' integral over the region of $\Sigma$
outside the horizon,
i.e.,
$\Sigma$ is a space-like hypersurface with boundaries at spatial
infinity and on the horizon.

The black holes satisfy the horizon mass formula
\begin{equation}
\frac{D-3}{D-2} M_{\rm  H} = \frac{\kappa A_{\rm H}}{8 \pi G_D} + N \Omega
J_{\rm H} \ . \label{hor_mass_form}
\end{equation}
They further satisfy the Smarr-like mass formula \cite{GMT}, 
\begin{equation}
M = \frac{D-2}{D-3} \frac{\kappa A_{\rm H}}{8 \pi G_D} + \frac{D-2}{D-3} N \Omega
J  +  \Phi_{\rm H} Q + \frac{D-5}{D-3} {\tilde \lambda} I \ , \label{smarr_like}
\end{equation}
where $I$ denotes the integral
\begin{equation}
I = -\frac{1}{4 \pi G_D} \int_\Sigma dS_\sigma \chi^\nu F_{\nu \rho} J^{\rho \sigma} \
, \label{int_I}
\end{equation}
and
$J^{\rho \sigma}$ is defined by
\begin{equation}
J^{\rho \sigma} = - \epsilon^{\rho \sigma \mu_1 \mu_2 \mu_3 \dots \mu_{D-3}
  \mu_{D-2}} A_{\mu_1} F_{\mu_2 \mu_3} \dots F_{\mu_{D-3} \mu_{D-2}} \ . \label{current} 
\end{equation}

\subsection{Scaling symmetry}

The system of ODEs is invariant under the scaling tranformation
\begin{equation}
r_{\rm H} \to \gamma r_{\rm H} \ , \ \ \Omega \to \Omega/\gamma \ , \ \
{\tilde \lambda} \to \gamma^{N-2} {\tilde \lambda} \ , \ \ Q \to \gamma^{D-3}
Q \ , \ \ a_\vphi \to \gamma a_\vphi \ , \label{scaling}
\end{equation}
with $\gamma>0$.
We note that in two cases the scaling transformation does
not change the CS coupling constant, namely, in the case ${\tilde \lambda}=0$ for
arbitrary dimension $D$ (i.e., in pure EM theory),
and in the case $D=5$ for arbitrary ${\tilde \lambda}$. 
This is in accordance with the mass formula,
Eq.~(\ref{smarr_like}). In both cases, the mass formula
reduces to the standard Smarr formula, since the last term vanishes. 
Indeed, this is not accidental, but both
features rely on the fact that the CS coupling constant is dimensionless only
for $D=5$, unless it vanishes \cite{GMT}.

\section{Numerical Results}

Apart from the case $D=5$, ${\tilde \lambda}=\tilde \lambda_{\rm SG}=1/(2\sqrt{3})$
\cite{BLMPSV, Cvetic}, no charged rotating EMCS black hole solutions
with spherical horizon topology are known analytically. 
Here we address the problem of finding such solutions numerically,
and discuss their properties.

\subsection{Numerical procedure}

Owing to the existence of the first integral of the system of ODE's
\begin{equation}
\frac{r^{D-2} m^{N-2}}{f^{N-1}} \sqrt{\frac{m n}{f}} \left(\frac{d
      a_0}{dr} + \frac{\omega}{r} \frac{d a_\vphi}{dr} \right)
-{\tilde \lambda}{\hat \vepsilon}_D 2^{D-2} (N-1)! a_\vphi^N
= - \frac{4 \pi G_D}{A(S^{D-2})} Q \ , \label{first_integral}
\end{equation}
we eliminate $a_0$ from the equations,
replacing it in terms of the electric charge.
This leaves a system of one first order equation (for $n$) and four second order
equations (for $f$, $m$, $\omega$, and $a_\vphi$). 
In Eq.~(\ref{first_integral}),
${\hat \vepsilon}_D$ is just a sign, depending on the
dimension D, and given by the expression
\begin{equation}
{\hat \vepsilon}_D \equiv (-1)^{N (N+1)/2} = (-1)^{(D^2-1)/8} \ . \label{sign_D}
\end{equation}

For the numerical calculations we then introduce
the compactified radial coordinate
$\bar{r}= 1-r_{\rm H}/r$ \cite{kkrot}, and we take units such that $G_D=1$.
We employ a collocation method for boundary-value ordinary
differential equations, equipped with an adaptive mesh selection procedure
\cite{COLSYS}.
Typical mesh sizes include $10^3-10^4$ points.
The solutions have a relative accuracy of $10^{-8}$. 
The set of numerical parameters to be fixed for a particular solution is
$\{r_{\rm H}, \Omega, Q, {\tilde \lambda}\}$. 
By varying these parameters we generate families of EMCS black holes. 

The scaling symmetry Eq.~(\ref{scaling}) leads us to consider 
$D=5$ black holes and black holes in $D>5$ dimensions separately. 
Since odd-dimensional EM black holes were investigated
previously \cite{EM_KNV}, we here concentrate on ${\tilde \lambda}
\neq 0$.

Moreover, since the action Eq.~(\ref{action}) is invariant under the
transformation
\begin{equation}
\{{\tilde \lambda}, Q\} \ \ \to \ \ \{(-1)^{N-1}{\tilde \lambda}, -Q\}
\label{discrete_sym} \ ,
\end{equation}
it is sufficient to consider just two of the four possible sign
combinations of $\{{\tilde \lambda}, Q\}$.


\boldmath
\subsection{$D=5$ EMCS black holes}
\unboldmath

For convenience we redefine the CS coupling constant,
\begin{equation}
\lambda=2\sqrt{3}\ {\tilde \lambda} \label{new_lambda} \ ,
\end{equation}
for $D=5$. Additionally, owing to Eq.~(\ref{discrete_sym}), without loss of
generality we may choose $\lambda \geq 0$.

$D=5$ EMCS black holes have been considered before 
for CS coupling $\lambda_{\rm SG}$ 
\cite{BLMPSV,Cvetic,GMT,surprise}
and for general coupling \cite{KN}.
To briefly recall some of their main features,
we exhibit
in Fig.~1a the domain of existence of these black holes.
Here the scaled angular momentum $|J|/M^{3/2}$ of extremal EMCS black holes
is shown versus the scaled charge $Q/M$ 
\footnote{
The scaling symmetry affects the physical quantities as follows,
$M \to \gamma^{D-3} M$,
$J \to \gamma^{D-2} J$,
$\mu_{\rm mag} \to  \gamma^{D-2}\mu_{\rm mag}$,
$g \to g$,
$\kappa \to \kappa/\gamma$, etc.}
for four values of $\lambda$:
the pure EM value $\lambda_{\rm EM}=0$,
the supergravity value $\lambda_{\rm SG}=1$,
the second critical value $\lambda_{\rm cr}=2$, and 
a value beyond this critical value, $\lambda=3$.
For a given value of $\lambda$,
black holes exist only in the regions bounded by the
$J=0$-axis and by the respective curves.
For fixed finite $\lambda$,
there is an explicit asymmetry between solutions with
positive and negative electric charge.
The properties of EMCS black holes with
positive $Q$ are similar to those of EM black holes \cite{EM_KNV},
whereas for EMCS black holes with negative $Q$ surprising features are present.
\begin{figure}[t!]
\parbox{\textwidth}{
\centerline{
\mbox{
\hspace{0.0cm} Fig.~1a \hspace{-2.0cm}
\epsfysize=6.0cm \epsffile{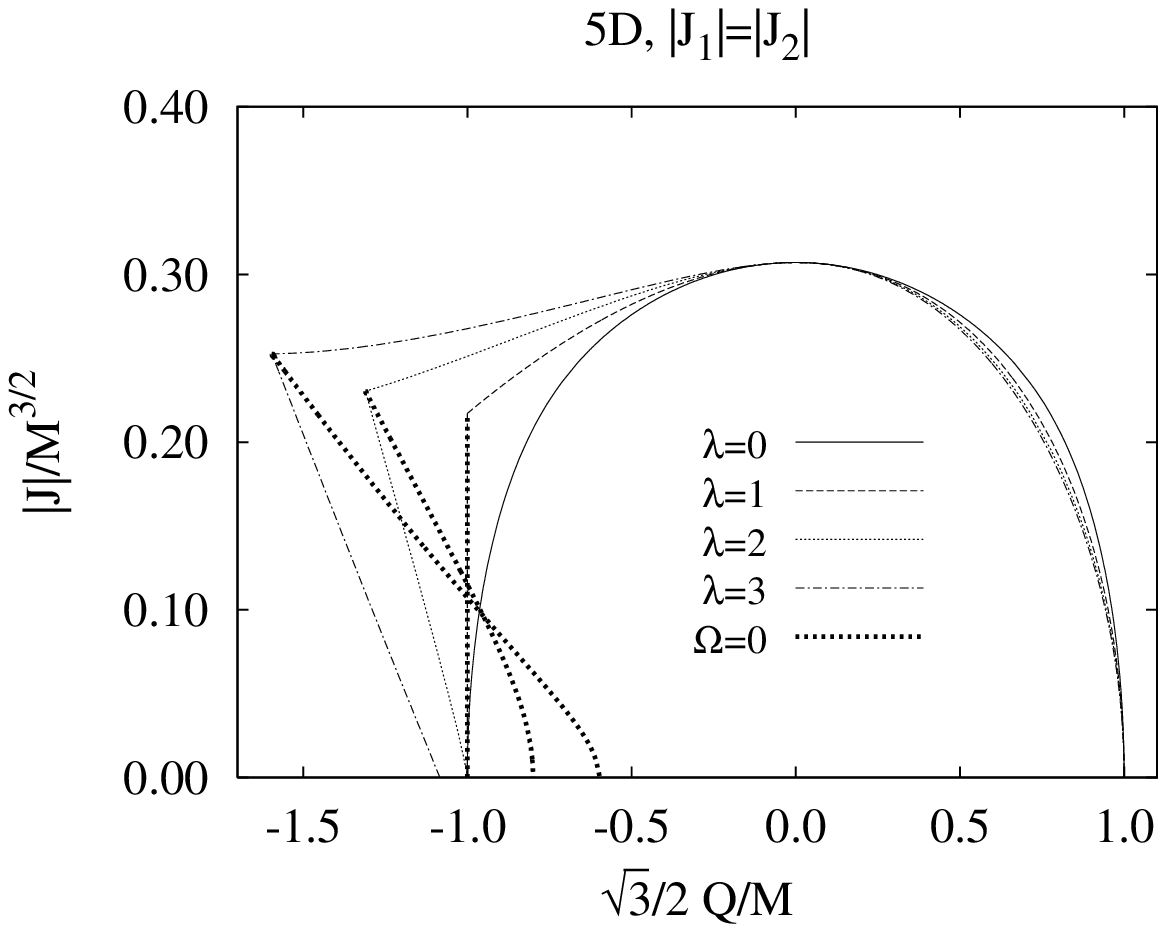} } \hspace{-1.cm}
\mbox{
\hspace{1.0cm} Fig.~1b \hspace{-2.0cm}
\epsfysize=6.0cm \epsffile{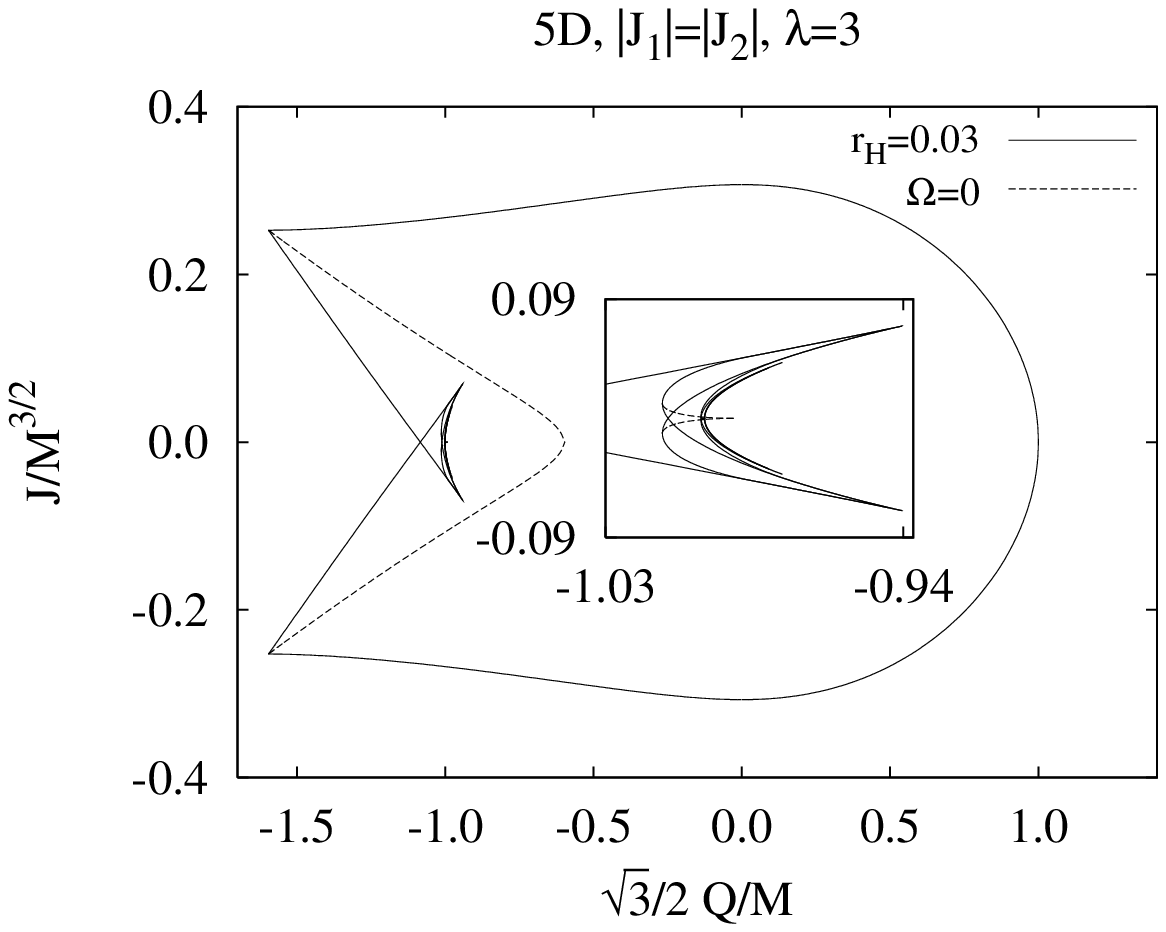} }
}}\vspace{0.5cm}
{\bf Fig.~1} \small
Scaled angular momentum $J/M^{3/2}$ versus scaled charge $Q/M$ for
$5D$ EMCS black holes.
a) Extremal and $\Omega=0$ solutions with CS coefficients $\lambda=0$, 1, 2, 3.
b) Almost extremal and $\Omega=0$ solutions with CS coefficient $\lambda=3$.
\vspace{0.2cm}
\end{figure}

The figure also shows stationary black holes with non-rotating horizon, i.e.,
black holes with horizon angular velocity $\Omega=0$ 
but finite total angular momentum.
Such solutions are present only for $\lambda \ge 1$.
For $\lambda=1$ they are extremal solutions,
and form the vertical part of the $Q<0$ borderline of the domain of existence.
For $\lambda>1$, however, they are (with the exception of endpoints)
non-extremal black holes, located in the interior of the respective 
domains of existence.
In fact, the $\Omega=0$ lines divide these domains into two parts.
The right part contains ordinary black holes,
where the horizon rotates in the same sense as the angular momentum,
while the left part contains black holes with unusual properties.
When $1<\lambda<2$, all black holes in this region are counterrotating,
i.e., their horizon rotates in the opposite sense to the angular momentum \cite{KN}.
When $\lambda>2$, further intriguing features appear.

As long as $1 < \lambda < 2$ only static $J=0$ solutions exist, and
consequently, the static extremal black holes are located on the borderline
of the domain of existence. (Recall that static black holes
are independent of $\lambda$.)
However, beyond $\lambda=2$,
a continuous set of rotating $J=0$ solutions appears \cite{KN},
whose existence relies on a special partition of the total
angular momentum $J$, where the horizon angular momentum 
$J_{\rm H}$ is equal and opposite to
the angular momentum $J_{\Sigma}$ in the Maxwell field outside the horizon
\footnote{
The numerical data indicate, that at $\lambda=2$
a continuous set of extremal rotating $J=0$ black holes with constant mass
is present \cite{KN}.
}.
The static extremal black holes then no longer mark the lower left
border of the domain of existence, which is now formed by
extremal rotating $J=0$ black holes,
as seen in Fig.~1a.

Moreover, beyond $\lambda=2$ the set of extremal solutions 
not only forms the boundary of the domain of existence,
but continues well within this domain,
until a static extremal black hole is reached
in a complicated pattern of bifurcating branches.
Insight into this set is gained in Fig.~1b, where (almost)
extremal solutions are exhibited for CS coefficient $\lambda=3$,
together with non-static $\Omega=0$ solutions.
Note, that all this new structure arises well within
the counterrotating region, in the vicinity of the static extremal black holes.

To explore the properties of $\lambda>2$ EMCS black holes further,
let us now consider non-extremal black holes.
We exhibit in Fig.~2 a set of solutions for $\lambda=3$,
possessing constant charge $Q=-10$ and 
constant (isotropic) horizon radius $r_{\rm H}=0.1$.
Fig.~2a and 2b show the total angular momentum $J$
and the horizon angular momentum $J_{\rm H}$, respectively,
as functions of the horizon angular velocity $\Omega$,
Fig.~2c and 2d show the corresponding masses $M$ and $M_{\rm H}$,
and Fig.~2e and 2f the gyromagnetic ratio $g$ and the horizon area $A_{\rm H}$.

\begin{figure}[p]
\parbox{\textwidth}{
\centerline{
\mbox{
\hspace{0.0cm} Fig.~2a \hspace{-2.0cm}
\epsfysize=6.0cm \epsffile{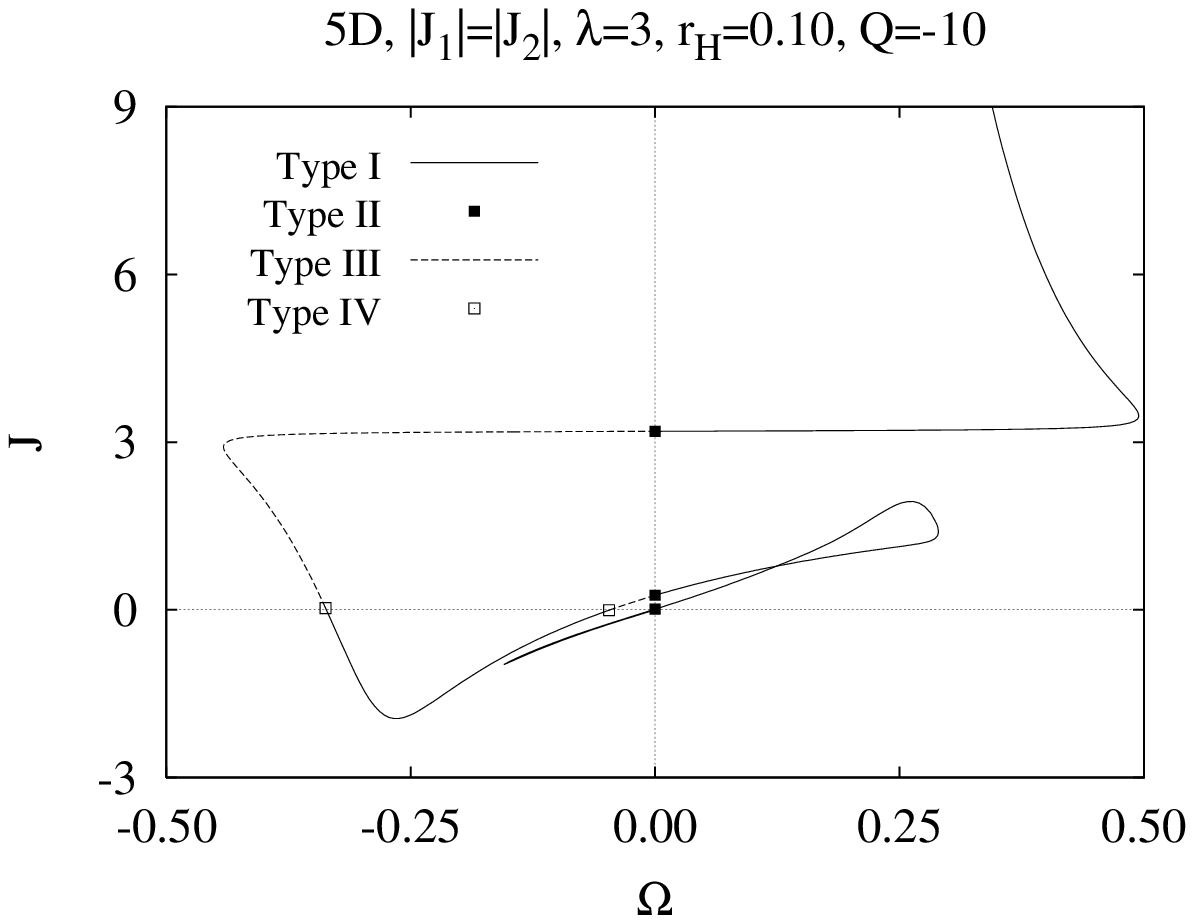} } \hspace{-1.cm}
\mbox{
\hspace{1.0cm} Fig.~2b \hspace{-2.0cm}
\epsfysize=6.0cm \epsffile{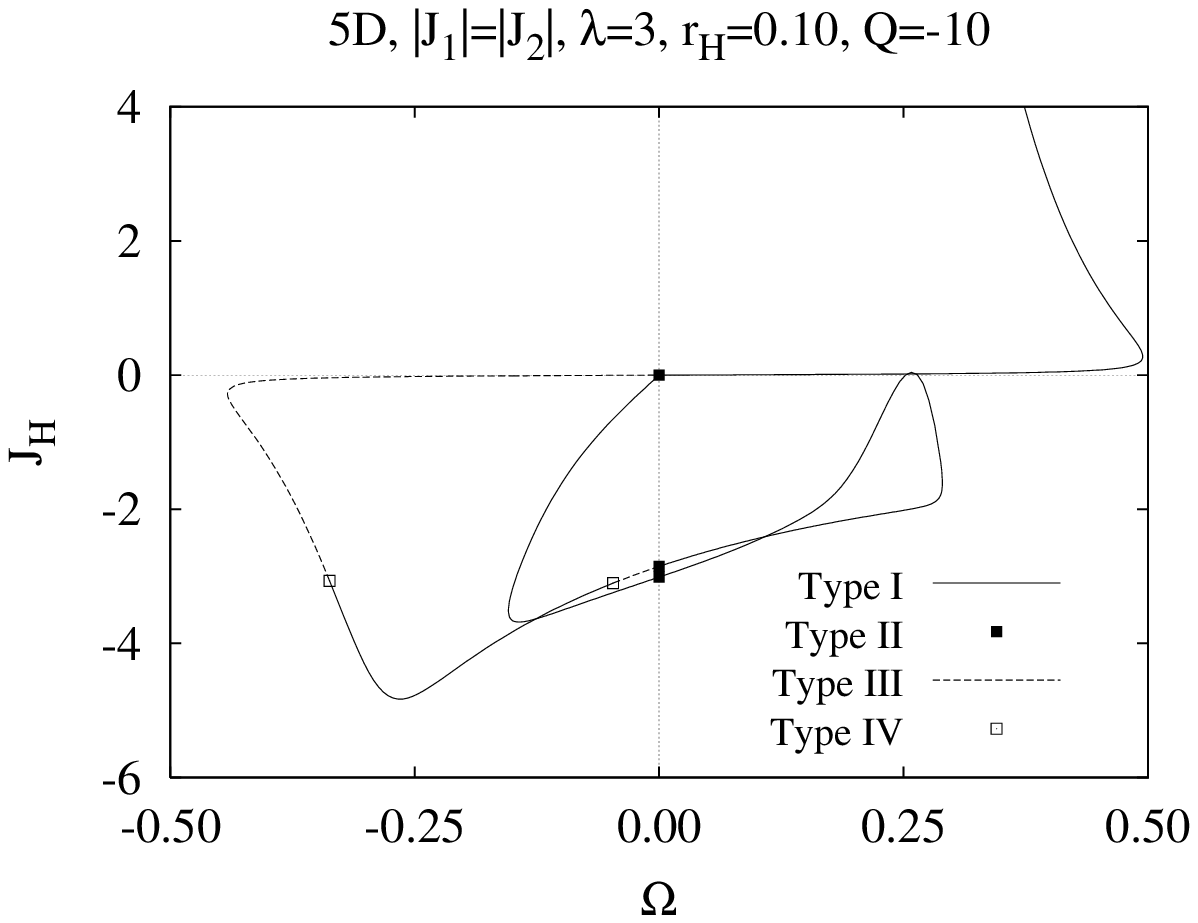} }
}\vspace{0.cm} }
\parbox{\textwidth}{
\centerline{
\mbox{
\hspace{0.0cm} Fig.~2c \hspace{-2.0cm}
\epsfysize=6.0cm \epsffile{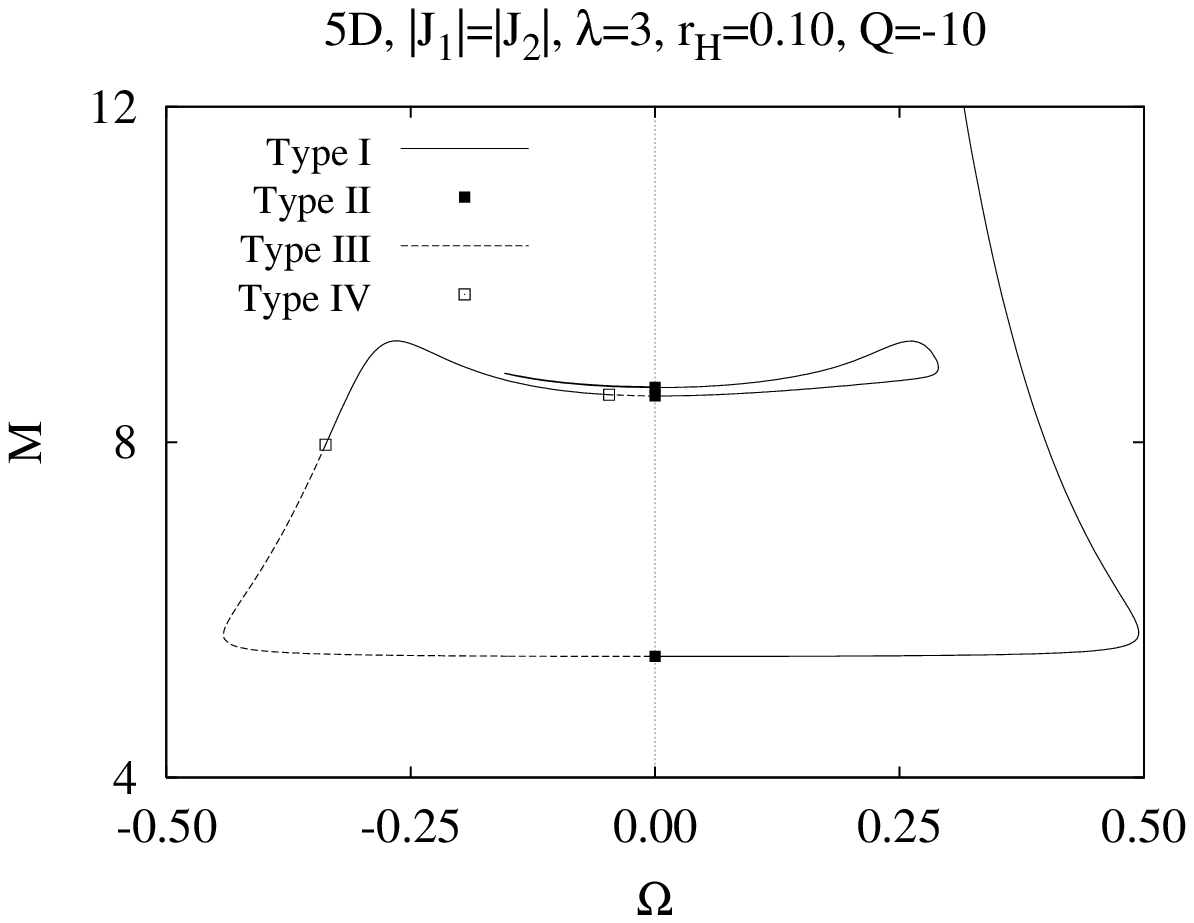} } \hspace{-1.cm}
\mbox{
\hspace{1.0cm} Fig.~2d \hspace{-2.0cm}
\epsfysize=6.0cm \epsffile{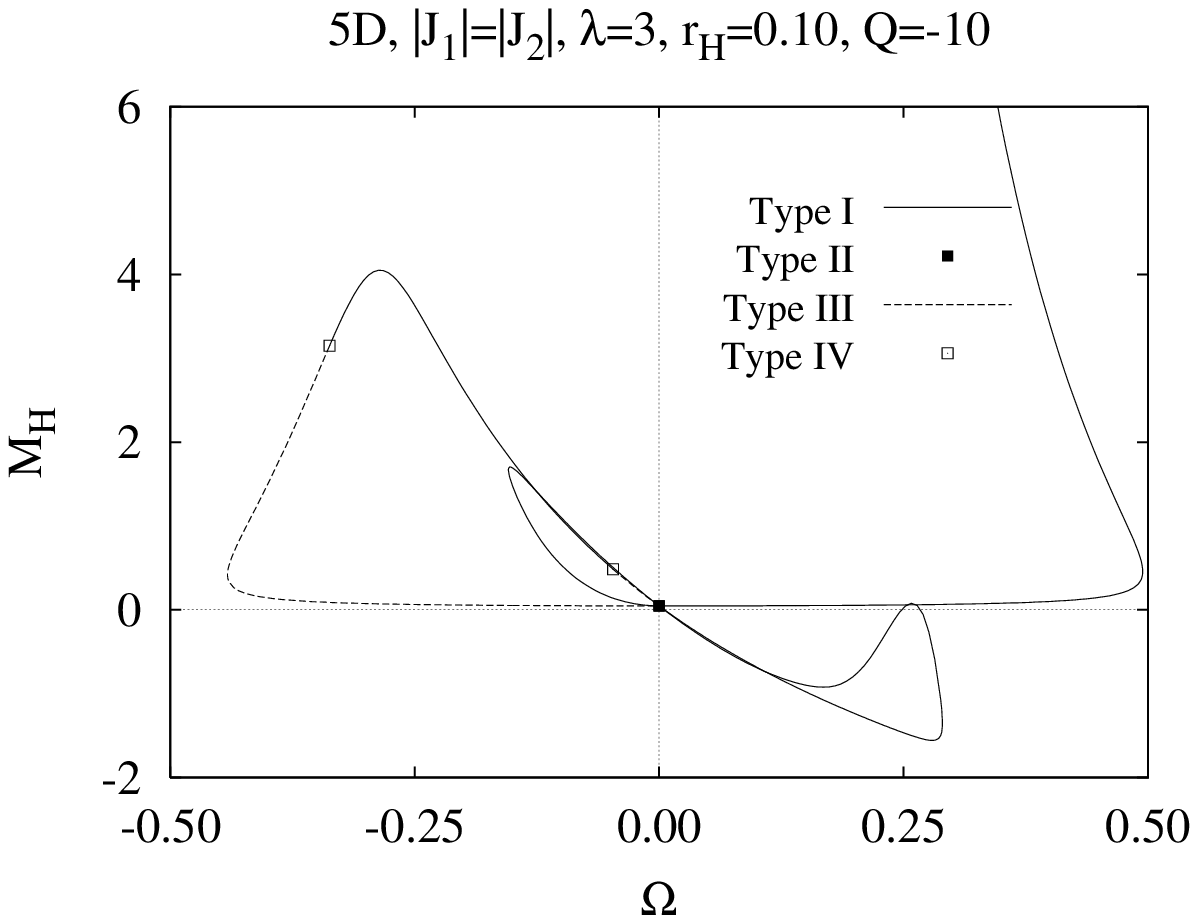} }
}\vspace{0.cm} }
\parbox{\textwidth}{
\centerline{
\mbox{
\hspace{0.0cm} Fig.~2e \hspace{-2.0cm}
\epsfysize=6.0cm \epsffile{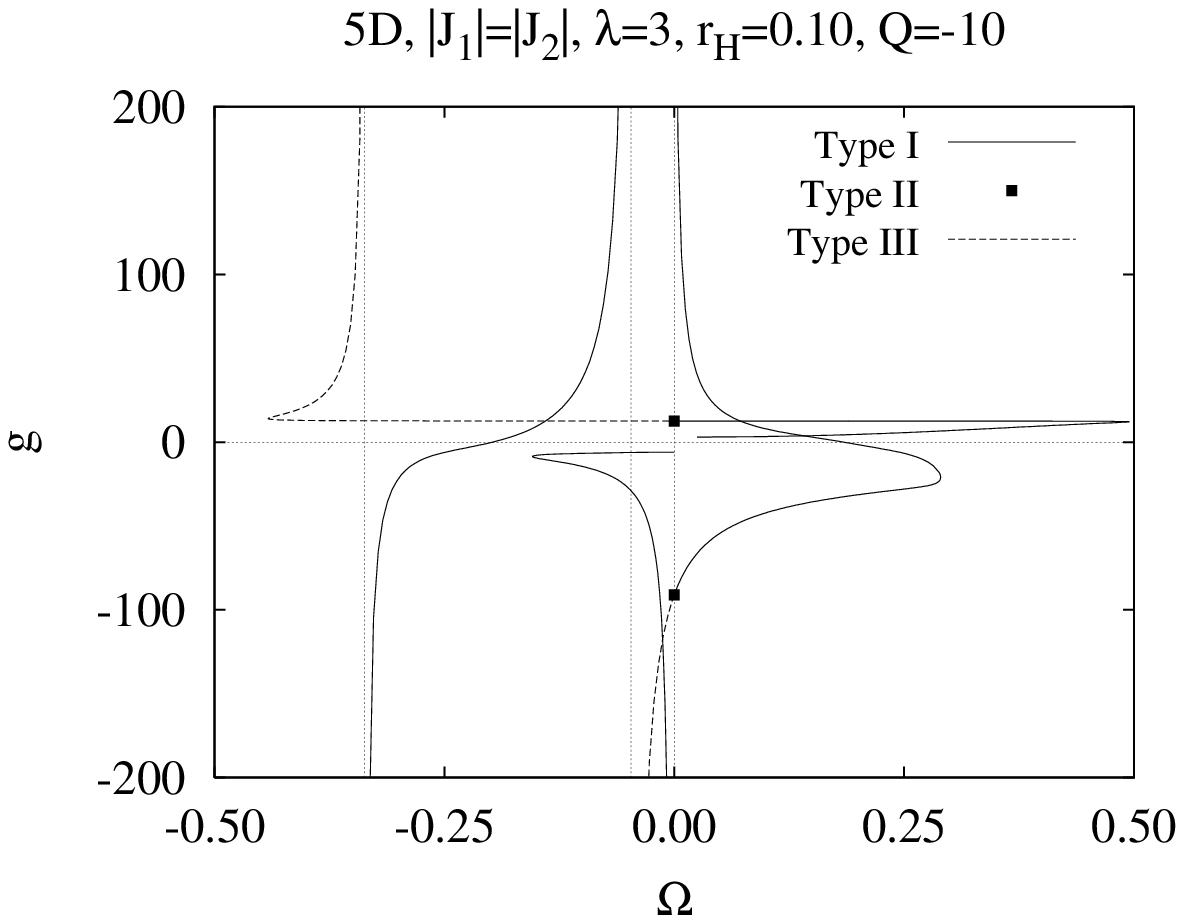} } \hspace{-1.cm}
\mbox{
\hspace{1.0cm} Fig.~2f \hspace{-2.0cm}
\epsfysize=6.0cm \epsffile{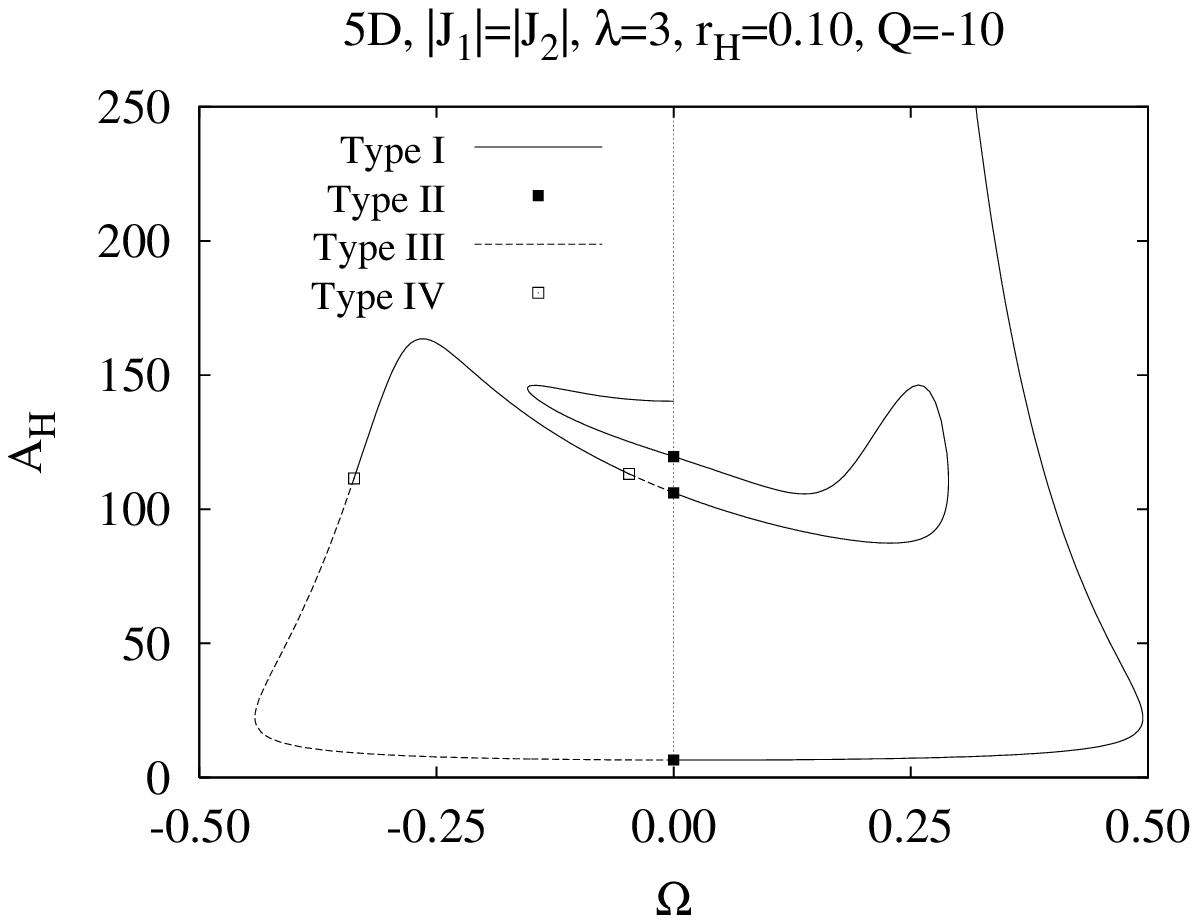} }
}}\vspace{0.5cm}
{\bf Fig.~2} \small
Properties of $5D$ non-extremal $\lambda=3$ EMCS black holes with
charge $Q=-10$ and horizon radius $r_{\rm H}=0.1$.
a) angular momentum $J$, b) horizon angular momentum $J_{\rm H}$,
c) mass $M$, d) horizon mass $M_{\rm H}$,
e) gyromagnetic ratio $g$, 
f) horizon area $A_{\rm H}$
versus horizon angular velocity $\Omega$.
\vspace{0.2cm}
\end{figure}

Fig.~2a exhibits the four types of rotating black holes, 
as classified by their total angular momentum $J$
and horizon angular velocity $\Omega$: Type I black
holes correspond to the corotating regime, i.e., $\Omega J \ge 0$, and
$\Omega=0$ if and only if $J=0$ (static). Type II black holes possess a static horizon
($\Omega=0$), although their total angular momentum does not vanish 
($J\neq 0$). 
Type III black holes are characterized by counterrotation, 
i.e., the horizon angular velocity
and the total angular momentum have opposite signs, $\Omega J < 0$. 
Type IV black holes, finally, possess a rotating horizon ($\Omega \neq
0$) but vanishing total angular momentum ($J=0$).  

As seen in Fig.~2b, the horizon angular momentum $J_{\rm H}$ of these
black holes need not have the same sign as the total angular momentum $J$,
and neither does the `bulk' angular momentum $J_{\Sigma}$.
In fact, as the horizon of the black hole is set into rotation,
angular momentum is stored in the Maxwell field both behind
and outside the horizon, yielding a rich variety of configurations.
Starting from the static solution, a corotating branch evolves,
along which $J_{\rm H}$ and $J_\Sigma$ have opposite signs.
After the first bifurcation $\Omega$ moves back towards zero
and so does $J$, but both $J_{\rm H}$ and $J_\Sigma$ remain finite,
retaining part of their built up angular momentum
and thus their memory of the path, like in a hysteresis.
This is important, since when moving $\Omega$ continuously 
back to and beyond zero, 
the total angular momentum follows and changes sign as well.
The horizon angular momentum, however, retains its sign.
Thus the product $\Omega J_{\rm H}$ turns negative 
and remains negative up to the next bifurcation
and still further, until $\Omega$ reaches zero again.

The observation, that solutions with $\Omega J_{\rm H}<0$
can be present, is crucial to 
explain the occurrence of black holes
with negative horizon mass, $M_{\rm H}<0$, exhibited in Fig.~2d.
The correlation between $\Omega J_{\rm H}<0$ and $M_{\rm H}<0$ black holes
is evident here. In fact,
the sets of $\Omega J_{\rm H}<0$ and $M_{\rm H}<0$ black holes almost coincide,
when $\kappa A_{\rm H}$ is small,
as seen from the horizon mass formula Eq.~(\ref{hor_mass_form}).
The angular momentum stored in the Maxwell field behind the horizon
is thus responsible for the negative horizon mass of the black holes.
The total mass is always positive, however, as seen in Fig.~2c.

The correlation between $\Omega J_{\rm H}<0$ and $M_{\rm H}<0$
is further demonstrated in Fig.~3, where we exhibit
$J_{\rm H}$ and $M_{\rm H}$ for black holes with
smaller as well as larger (isotropic) horizon radii $r_{\rm H}$.
\begin{figure}[h!]
\parbox{\textwidth}{
\centerline{
\mbox{
\hspace{0.0cm} Fig.~3a \hspace{-2.0cm}
\epsfysize=6.0cm \epsffile{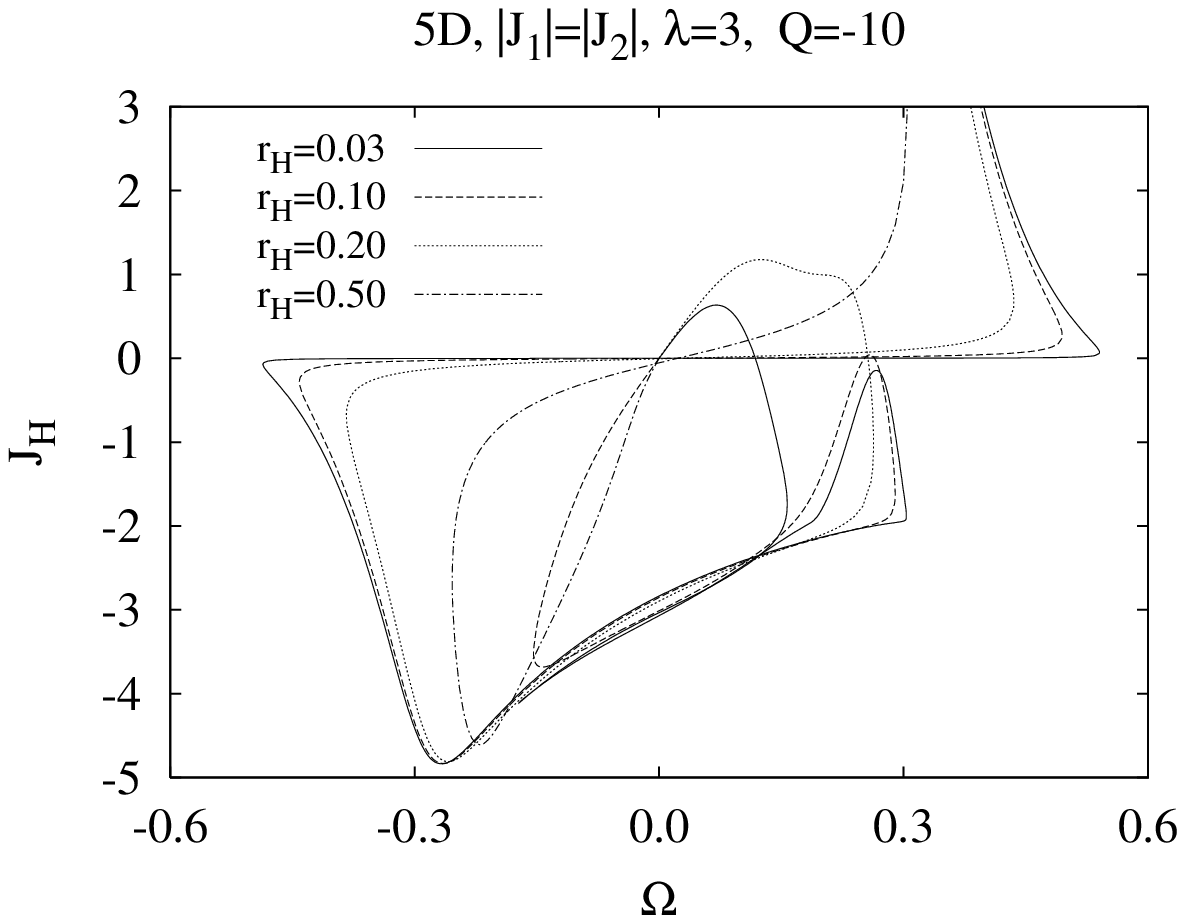} } \hspace{-1.cm}
\mbox{
\hspace{1.0cm} Fig.~3b \hspace{-2.0cm}
\epsfysize=6.0cm \epsffile{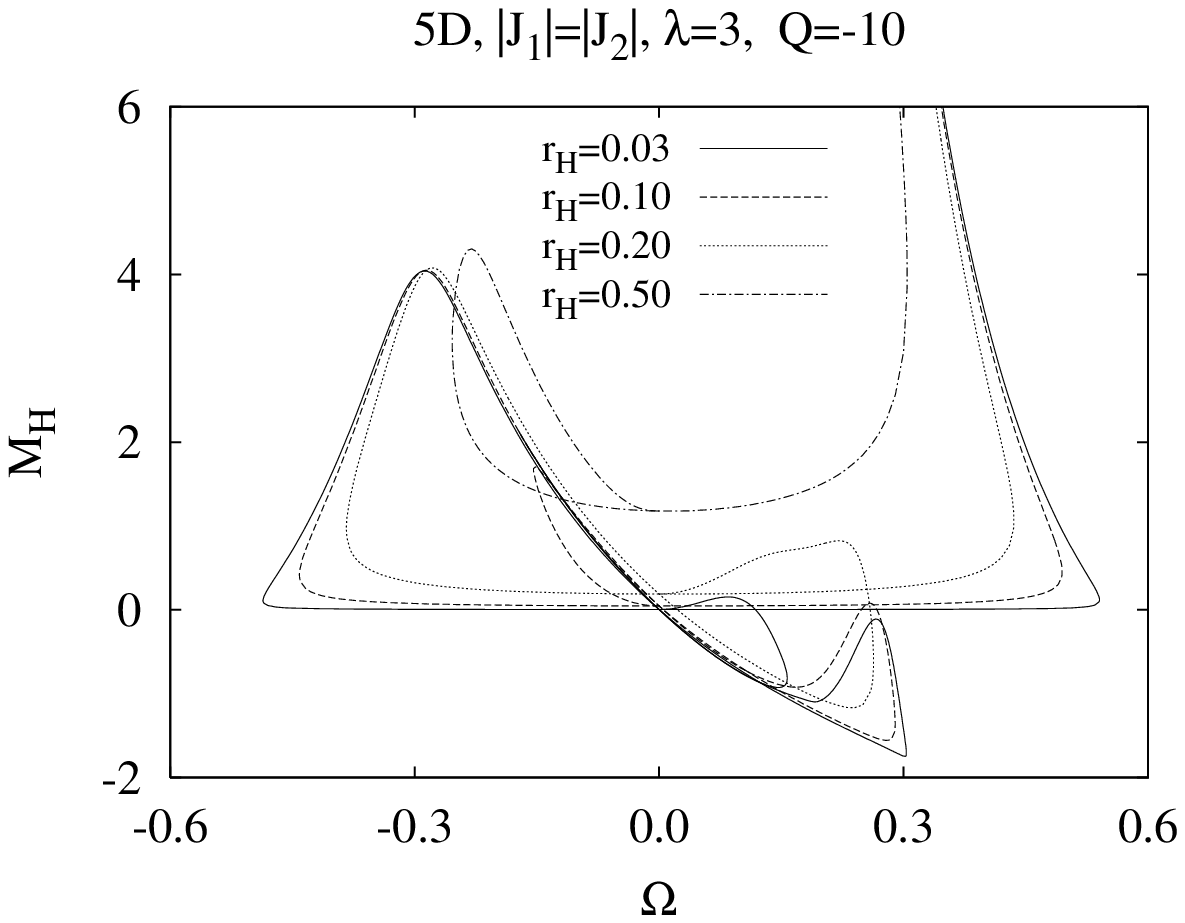} }
}}\vspace{0.5cm}
{\bf Fig.~3} \small
Properties of $5D$ non-extremal $\lambda=3$ EMCS black holes with
charge $Q=-10$ and horizon radii $r_{\rm H}=0.03$, 0.1, 0.2 and 0.5.
a) horizon angular momentum $J_{\rm H}$,
b) horizon mass $M_{\rm H}$
versus horizon angular velocity $\Omega$.
\vspace{0.2cm}
\end{figure}
Whereas for larger values of $r_{\rm H}$ the set of negative horizon
mass black holes decreases, it increases for smaller values of $r_{\rm H}$.
In fact, as $r_{\rm H}$ decreases, more branches of solutions appear
in the vicinity of the static solution, giving rise to more branches
of negative horizon mass black holes.

Another interesting feature of these charged rotating EMCS black holes 
is their gyromagnetic ratio $g$, exhibited in Fig.~2e.
The gyromagnetic ratio is unbounded, reaching any real value
including zero.
The main consequence of this is that, contrary to pure EM theory, 
a vanishing total angular momentum does not readily
imply a vanishing magnetic moment and viceversa.

\boldmath
\subsection{$D>5$ EMCS black holes}
\unboldmath

Unlike in $D=5$, the CS coupling constant is
dimensionful when $D>5$, and consequently 
changes under scaling transformations Eq.~(\ref{scaling}), unless
${\tilde \lambda=0}$. 
Thus any feature present for a certain
non-vanishing value of ${\tilde \lambda}$ 
will be present for any other non-vanishing value,
when the charge is scaled correspondingly.
To classify the critical behaviour of the solutions
we therefore consider classes of solutions
labelled by the scale invariant ratio $\lambda_Q$,
\begin{equation}
\lambda_Q = {\tilde \lambda}/|Q|^{(D-5)/2(D-3)}
\ . \label{lambdaQ} \end{equation}

Otherwise, black holes of $D>5$ EMCS theories
exhibit very similar properties to $D=5$ EMCS black holes.
Besides regions containing black holes with features
resembling those of EM black holes, there are regions
where black holes reside, possessing all those new features.
Assuming $Q<0$, the sign of ${\tilde \lambda}$ associated with
the appearance of these new types of black holes
coincides with $-{\hat \vepsilon}_D$ (while for $Q>0$ one needs to
employ Eq.~(\ref{discrete_sym})).

We exhibit in Fig.~4a sets of extremal $7D$ solutions,
choosing the same value of the charge in all sets,
while varying $\tilde \lambda$.
In $7D$, the first critical value is $\lambda_Q = -0.0414$.
Here the first $\Omega=0$ solution appears.
Counterrotating black holes then exist only for values of the ratio
$\lambda_Q$ above this critical value.
The second critical value characterizes the limit above which
non-static $J=0$ solutions are present, 
and where solutions are no longer uniquely specified by their global charges.
In $7D$ this value is $\lambda_Q = -0.2101$.
We note, that the sets of $\Omega=0$ solutions, also exhibited in the figure,
always connect extremal solutions, 
whose mass assumes an extremal value as well, 
in accordance with the first law.
At the second critical value of $\lambda_Q$, the lower endpoint 
of the $\Omega=0$ set of solutions
precisely reaches a static extremal black hole.
\begin{figure}[h!]
\parbox{\textwidth}{
\centerline{
\mbox{
\hspace{0.0cm} Fig.~4a \hspace{-2.0cm}
\epsfysize=6.0cm \epsffile{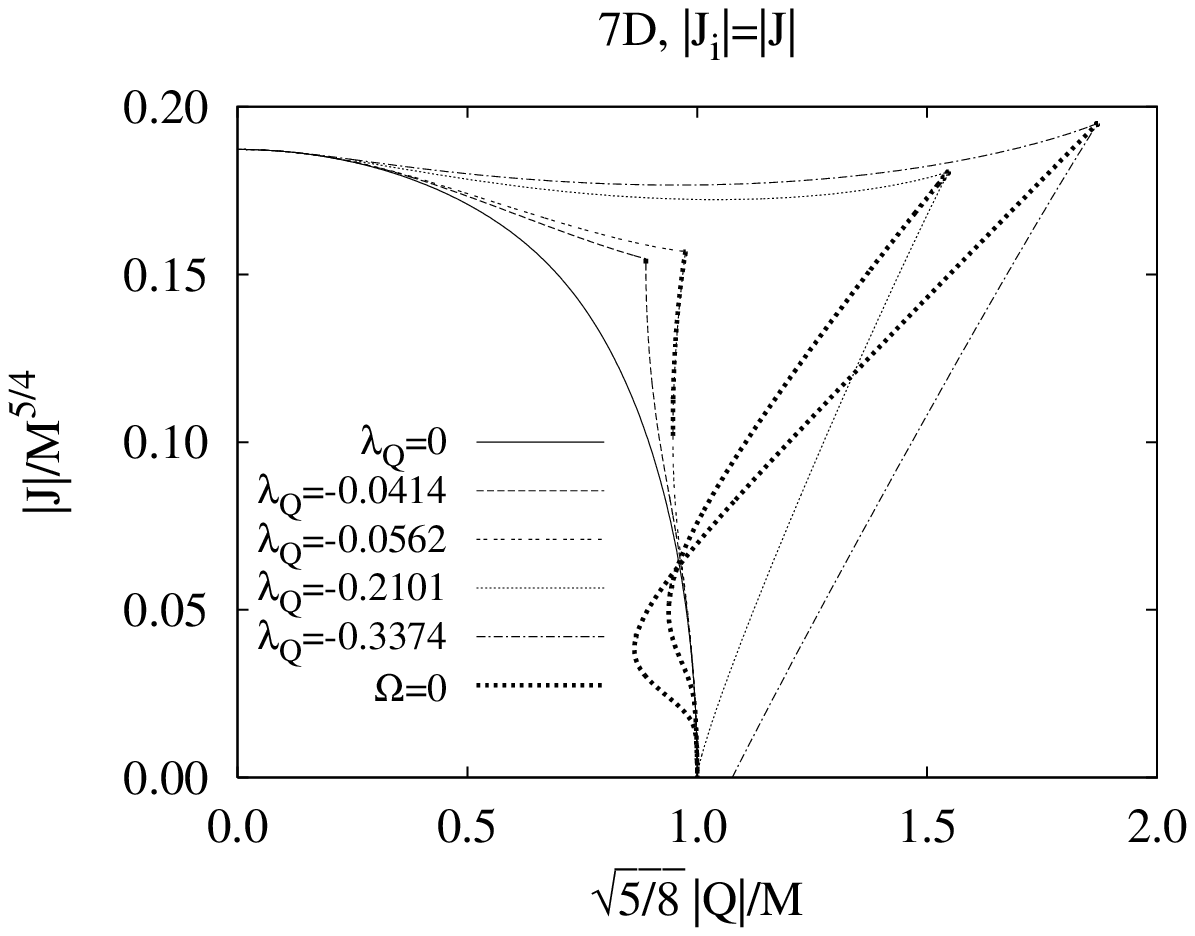} } \hspace{-1.cm}
\mbox{
\hspace{1.0cm} Fig.~4b \hspace{-2.0cm}
\epsfysize=6.0cm \epsffile{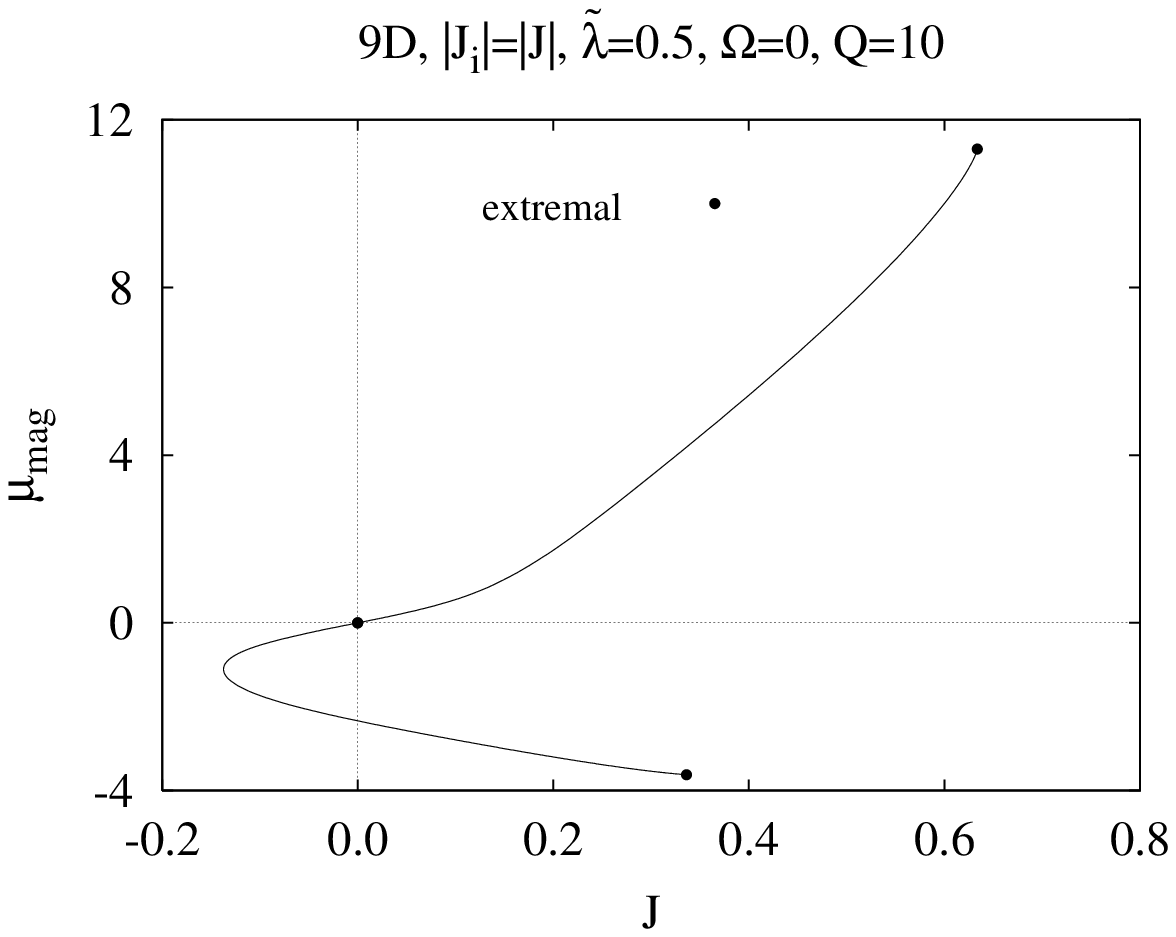} }
}}\vspace{0.5cm}
{\bf Fig.~4} \small
a) Scaled angular momentum $J/M^{5/4}$ versus scaled charge $Q/M$ for
(almost) extremal $7D$ EMCS black holes and $\Omega=0$ solutions
for $\lambda_Q=0$, -0.0414, -0.0562, -0.2101, -0.3374
and charge $Q=10$.
b) Magnetic moment $\mu_{\rm mag}$ versus total angular momentum $J$
for $9D$ EMCS black holes with $\Omega=0$ for $\tilde \lambda=0.5$
and charge $Q=10$.
\vspace{0.2cm}
\end{figure}

In Fig.~5 we demonstrate that the four types of rotating black holes,
as classified by their total angular momentum $J$
and horizon angular velocity $\Omega$, are also present for $D>5$.
Interestingly, in $D=9$, beside
I. corotating black holes, 
II. non-static $\Omega=0$ black holes,
III. counterrotating black holes,
IV. non-static $J=0$ black holes,
a further type of black holes appears:
V. non-static $\Omega=J=0$ black holes.
This is seen in Fig.~4b, where the magnetic moment is exhibited
for a set of $9D$ $\Omega=0$ black holes. Clearly, there appears a
non-static $\Omega=J=0$ solution with finite magnetic moment.
We further illustrate in Fig.~5 that these EMCS
black holes are no longer uniquely characterized by their global charges,
i.e., the uniqueness conjecture does not hold in general
for $D \ge 5$ EMCS
stationary black holes with horizons of spherical topology
\footnote{
Previous counterexamples involved black rings \cite{blackrings}.}.
\begin{figure}[h!]
\parbox{\textwidth}{
\centerline{
\mbox{
\hspace{0.0cm} Fig.~5a \hspace{-2.0cm}
\epsfysize=6.0cm \epsffile{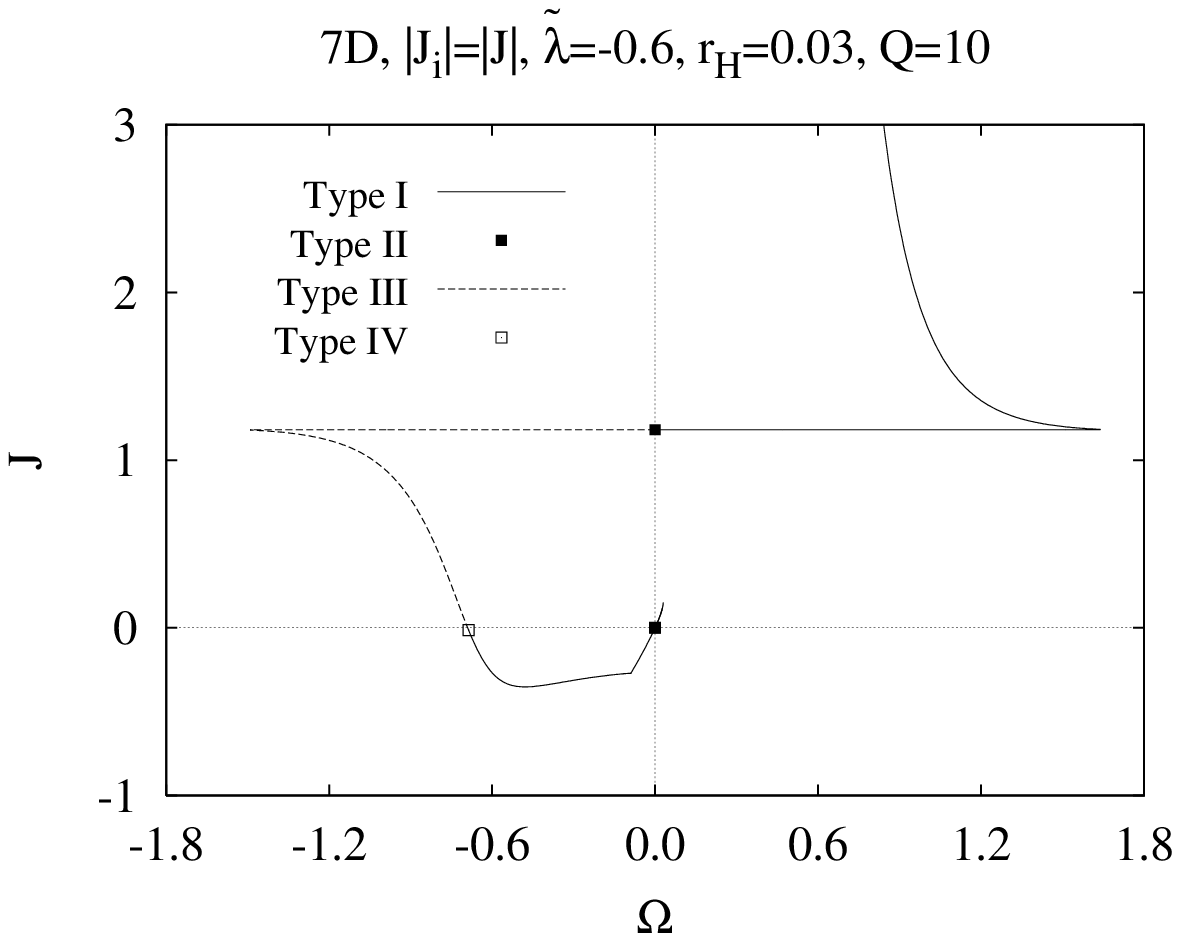} } \hspace{-1.cm}
\mbox{
\hspace{1.0cm} Fig.~5b \hspace{-2.0cm}
\epsfysize=6.0cm \epsffile{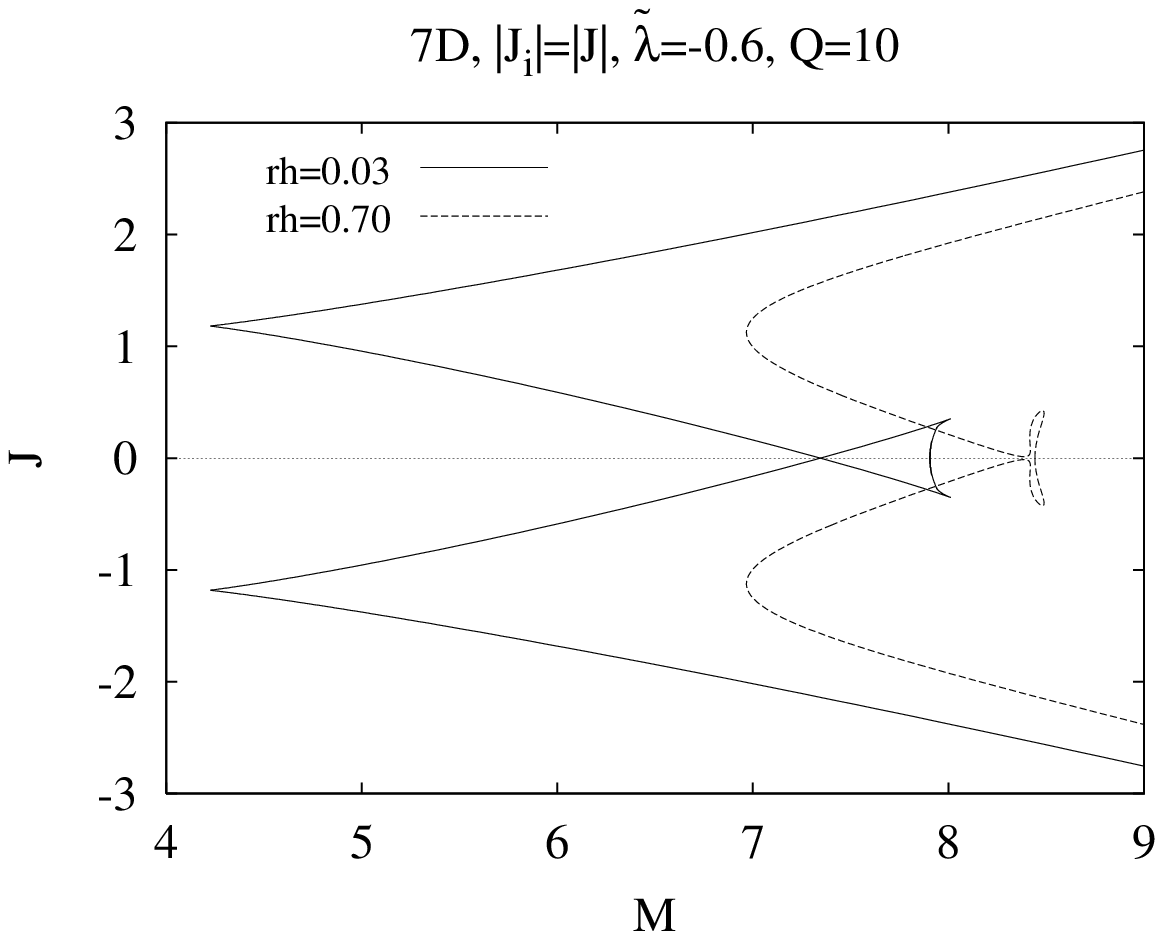} }
}}\vspace{0.5cm}
{\bf Fig.~5} \small
Angular momentum $J$ of $7D$ non-extremal ${\tilde \lambda}=-0.6$ EMCS black holes with
charge $Q=10$ 
a) versus horizon angular velocity $\Omega$
for horizon radius $r_{\rm H}=0.03$,
b) versus mass $M$ for horizon radii $r_{\rm H}=0.03$, 0.7.
\vspace{0.2cm}
\end{figure}

We illustrate in Fig.~6 that $D>5$ EMCS black holes can also possess
negative horizon mass $M_{\rm H}$, by exhibiting sets of $D=7$ and
$D=9$ black holes. As in $D=5$ dimensions,
the angular momentum stored in the Maxwell field behind the horizon
is responsible for this phenomenon.
Finally, the gyromagnetic ratio $g$ is unbounded as well for 
$D>5$ EMCS black holes, reducing the correlation
between angular momentum and magnetic moment. 
\begin{figure}[h!]
\parbox{\textwidth}{
\centerline{
\mbox{
\hspace{0.0cm} Fig.~6a \hspace{-2.0cm}
\epsfysize=6.0cm \epsffile{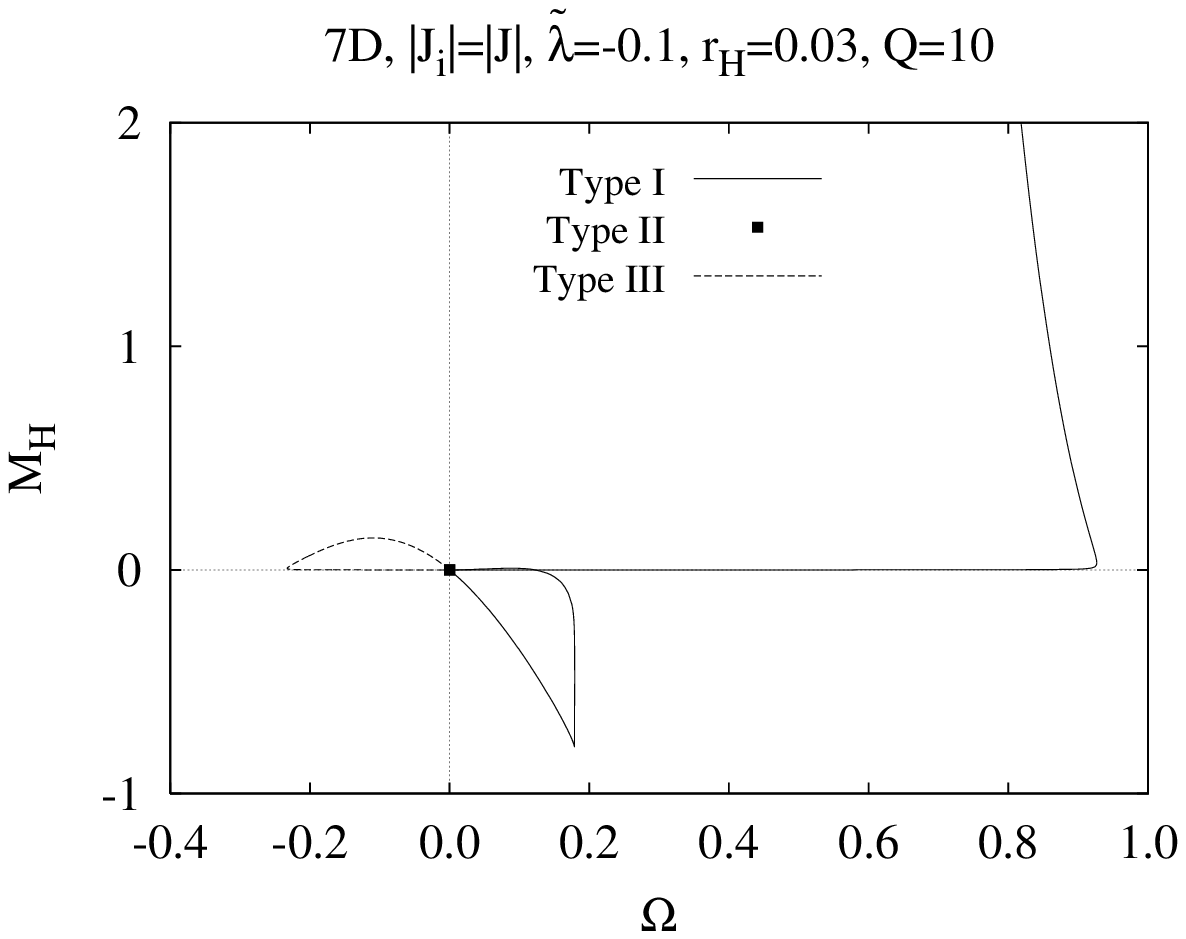} } \hspace{-1.cm}
\mbox{
\hspace{1.0cm} Fig.~6b \hspace{-2.0cm}
\epsfysize=6.0cm \epsffile{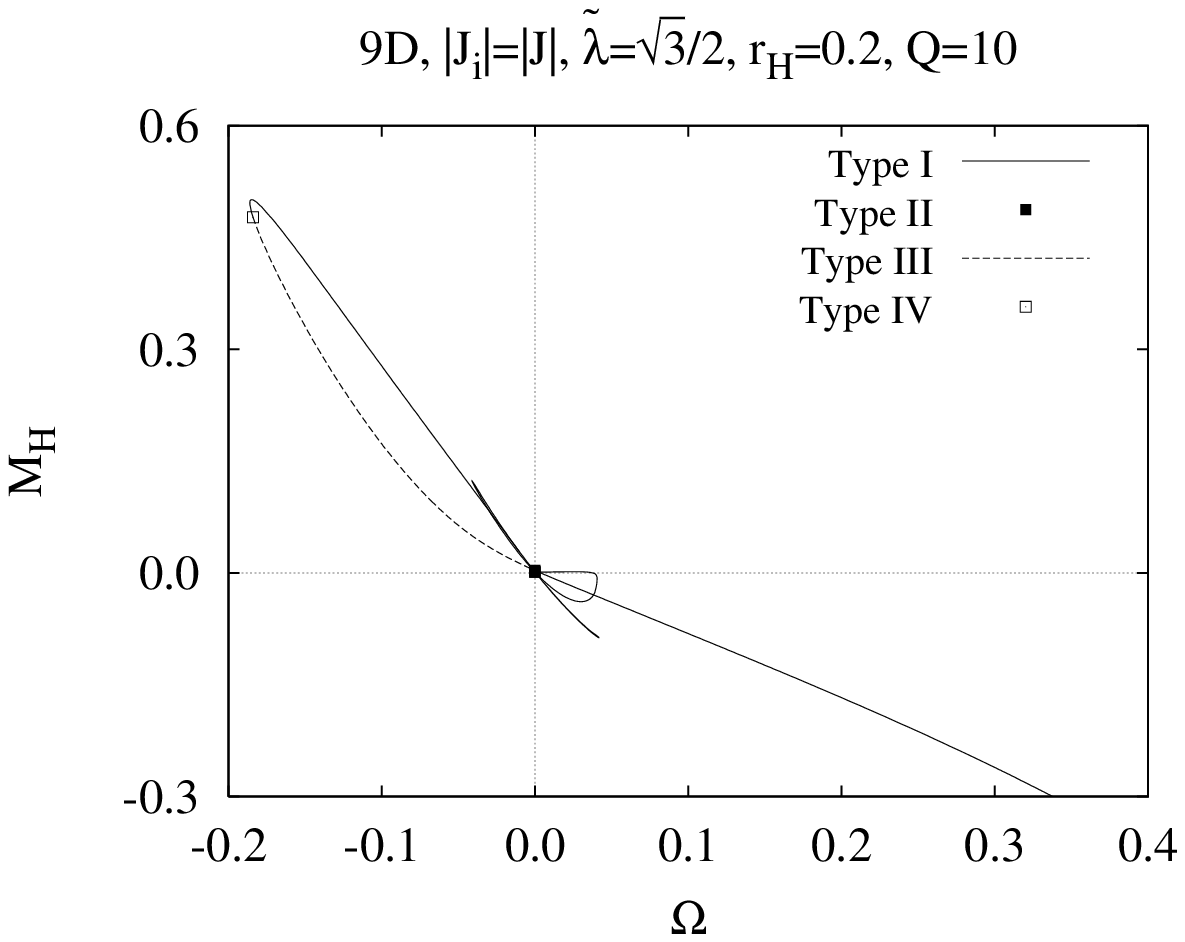} }
}}\vspace{0.5cm}
{\bf Fig.~6} \small
Horizon mass $M_{\rm H}$ of non-extremal EMCS black holes with
charge $Q=10$ versus horizon angular velocity $\Omega$
a) for $7D$, horizon radius $r_{\rm H}=0.03$, ${\tilde \lambda=-0.1}$,
b) for $9D$, horizon radius $r_{\rm H}=0.2$, ${\tilde \lambda}=\sqrt{3}/2$.
\vspace{0.2cm}
\end{figure}

\section{Conclusions}

We have considered charged rotating black holes 
of EMCS theory in odd dimensions,
which are asymptotically flat
and possess a regular horizon of spherical topology.
When all their angular momenta have equal-magnitude,
a system of 5 ordinary differential
equations results \cite{EM_KNV},
which we have solved numerically.

EMCS black holes exhibit remarkable features, not
present for EM black holes.
Classifying the EMCS black holes by their total angular momentum $J$
and horizon angular velocity $\Omega$, we find several types of black holes:
I. corotating black holes, i.e., $\Omega J \ge 0$,
II. black holes with static horizon and 
non-vanishing total angular momentum, i.e., $\Omega=0$, $J\neq 0$,
III. counterrotating black holes, where the horizon angular velocity
and the total angular momentum have opposite signs, i.e., $\Omega J < 0$,
and IV. black holes with rotating horizon and
vanishing total angular momentum, i.e., $\Omega \neq 0$, $J=0$.
Furthermore, in $9D$ type V. black holes appear: non-static black holes
with static horizon and vanishing total angular momentum, i.e., $\Omega=J=0$.

As the horizon of static EMCS black holes is set into rotation,
angular momentum is stored in the Maxwell field both behind
and outside the horizon. 
Following paths through configuration space, 
the horizon angular momentum $J_{\rm H}$ 
and the `bulk' angular momentum $J_\Sigma$
can retain part of the angular momentum built up in the Maxwell field,
even when the horizon angular velocity vanishes again.
Thus they retain the memory of the path, like a hysteresis.
Consequently, solutions with $\Omega J_{\rm H}<0$
appear, which possess a negative horizon mass, $M_{\rm H}<0$,
as long as $\kappa A_{\rm H}$ is sufficiently small.
The angular momentum stored in the Maxwell field behind the horizon
is therefore responsible for the negative horizon mass of these black holes.
Their total mass is always positive, however \footnote{
The occurrence of a negative horizon mass has
been also reported recently in the context of $4D$ black holes
surrounded by perfect fluid rings \cite{ansorg}.}.

Moreover, EMCS black holes are no longer uniquely characterized 
by their global charges,
i.e., the uniqueness conjecture does not hold in general
for $D \ge 5$ EMCS
stationary black holes with horizons of spherical topology,
and the gyromagnetic ratio of these black holes
may take any real value, including zero.

{\bf Acknowledgement}

FNL gratefully acknowledges Ministerio de Educaci\'on y Ciencia for
support under grant EX2005-0078.

\end{document}